# High polarization lines of the second solar spectrum of the Solar limb

*Jean-Marie Malherbe (emeritus astronomer)*
Observatoire de Paris, PSL Research University, LIRA, France
Email: Jean-Marie.Malherbe@obspm.fr; ORCID: https://orcid.org/0000-0002-4180-3729
2 May 2025

**ABSTRACT**

We present a dataset of high resolution spectra of the Sun of many strongly polarized lines belonging to the second solar spectrum, i.e. the spectrum near the limb in linear polarization (scattering polarization). These solar spectra were obtained in full Stokes polarimetry (I, Q/I, U/I, V/I) in the quiet Sun at various distances from the limb, and at disk centre for comparison, with the ground based CNRS THEMIS telescope. Polarization rates Q/I up to 7% are obtained in CaI 4227 Å line at $\mu = \cos\theta = 0$, while 2% is reached in SrI 4607 Å line and 1.4% in BaII 4554 Å. The spectra shown here are freely available in FITS format to the research community.

**INTRODUCTION**

The second solar spectrum is the linearly polarized spectrum observed at the limb. It is formed by coherent scattering due to the anisotropy of the radiation field. The polarization rate Q/I is generally small (< 1%) and decreases in the presence of magnetic fields (the Hanle effect). The first survey of the second solar spectrum was performed at Kitt Peak by Stenflo *et al.* (1983a, b). It was then followed by a more sensitive survey by Gandorfer (2000, 2002) using the ZIMPOL polarimeter. The second solar spectrum requires polarization free telescopes to be best recorded and opened a new field of research in solar physics, such as diagnostics of spatially unresolved, turbulent and weak, magnetic fields. Stenflo (2004) noticed from observations made at two different periods (1995, close to the minimum activity, and 2000) that the proportion of absorption-like and emission-like spectral lines varies, which suggests the varying influence of hidden magnetic fields (more Hanle depolarization at maximum). The analysis of long-term fluctuations of the second solar spectrum could therefore be useful to monitor the magnetic cycle.

**I - THE EXPERIMENT SETUP**

**I – 1- Dual beam configuration for Stokes vector measurements**

The configuration of Figure 1 was used. The birefringent separator is equivalent to 2 polarizers of Ox and Oy axes, i.e. the two outgoing beams S1 and S2 are linearly polarized in the two directions Ox and Oy. But unlike two polarizers, $S_1$ and $S_2$ coexist simultaneously and are spatially shifted. The optical axis 2 of the separator is inclined with respect to the xOy plane by an angle φ. There is a maximum deviation $\varphi_m$ such that:

**cos 2$\varphi_m$ = ($n_e^2 - n_o^2$) / ($n_e^2 + n_o^2$)**
$\varphi_m = 48°$ (with $n_o = 1.658$ and $n_e = 1.486$ for calcite). The shift s between $S_1$ and $S_2$ is given by:

**s / e = sin(2φ) / ( cos(2φ) + ($n_o^2 + n_e^2$) / ( $n_o^2 - n_e^2$) )**
at the deviation maximum, we have: **s = e ($n_o^2 - n_e^2$) / (2 $n_o n_e$)**
and one gets **s = 0.11 e** for calcite (in average, as it depends on wavelength)

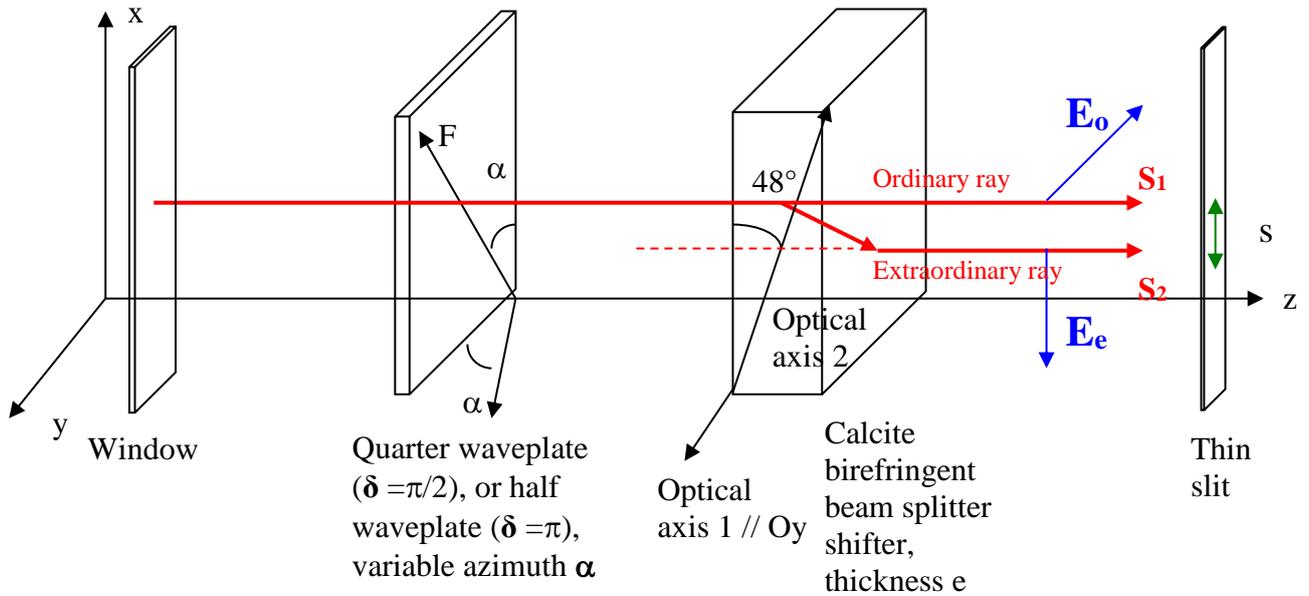

*Figure 1 : principle of the THEMIS polarimeter (courtesy OP)*

The calcite separator has chromatism (because $n_o$ and $n_e$ vary according to the wavelength), as shown in Figure 2. For that reason, when the grid is used (Figure 4), the separation width between the strips depends on the wavelength.

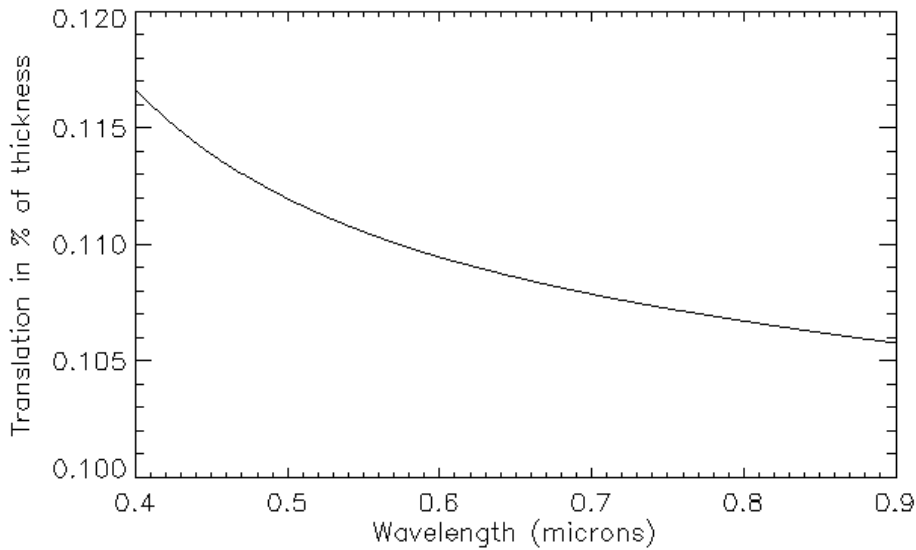

*Figure 2 : shift s / e as a function of wavelength for $\varphi = 48.15°$*

## I – 2 - full stokes polarimetry with 2 rotating quarter waveplates and a birefringent beam splitter shifter

If $\alpha$ and $\beta$ are the azimuths of the two waveplates, it can be shown that we measure the two quantities $S_1$ and $S_2$ simultaneously:

$S_1 = \frac{1}{2} [ I_{in} + \{ Q_{in} ( \cos(2(\beta-\alpha)) \cos(2\alpha) \cos(2\beta) - \sin(2\alpha) \sin(2\beta) ) + U_{in} ( \cos(2(\beta-\alpha)) \sin(2\alpha) \cos(2\beta) + \cos(2\alpha) \sin(2\beta) ) + V_{in} \sin(2(\beta-\alpha)) \cos(2\beta) \} ]$

$S_2 = \frac{1}{2} [ I_{in} - \{ Q_{in} ( \cos(2(\beta-\alpha)) \cos(2\alpha) \cos(2\beta) - \sin(2\alpha) \sin(2\beta) ) + U_{in} ( \cos(2(\beta-\alpha)) \sin(2\alpha) \cos(2\beta) + \cos(2\alpha) \sin(2\beta) ) + V_{in} \sin(2(\beta-\alpha)) \cos(2\beta) \} ]$

According to α, β values, one gets the couples $S_1$, $S_2$:

$\alpha = \beta = 0$,          $S_1 = ½ [ I_{in} + Q_{in} ]$ et $S_2 = ½ [ I_{in} - Q_{in} ]$
$\alpha = \beta = \pi/4$,      $S'_1 = ½ [ I_{in} - Q_{in} ]$ et $S'_2 = ½ [ I_{in} - Q_{in} ]$ beam exchange for Q
$\alpha = \beta = \pi/8$,      $S_1 = ½ [ I_{in} + U_{in} ]$ et $S_2 = ½ [ I_{in} - U_{in} ]$
$\alpha = \beta = 3\pi/8$,     $S'_1 = ½ [ I_{in} - U_{in} ]$ et $S'_2 = ½ [ I_{in} + U_{in} ]$ beam exchange for U
$\alpha = \pi/4, \beta = 0$,   $S_1 = ½ [ I_{in} - V_{in} ]$ et $S_2 = ½ [ I_{in} + V_{in} ]$
$\alpha = \pi/4, \beta = \pi/2$, $S'_1 = ½ [ I_{in} + V_{in} ]$ et $S'_2 = ½ [ I_{in} - V_{in} ]$ beam exchange for V

The waveplate retardance δ is never exactly equal to π/2, the formula is $\delta = (2 \pi / \lambda) (n_e - n_o) e$ where e is the thickness and $(n_e - n_o)$ the difference of the ordinary and extraordinary refractive index. The waveplates made of a single material have strong chromatism. Two-material assemblies are preferred as shown by Figure 3. The assembly of MgF2/quartz of the THEMIS polarimeter does not exceed a departure of ±1° for a 90° retardance quarter waveplate in the interval [450-600] nm.

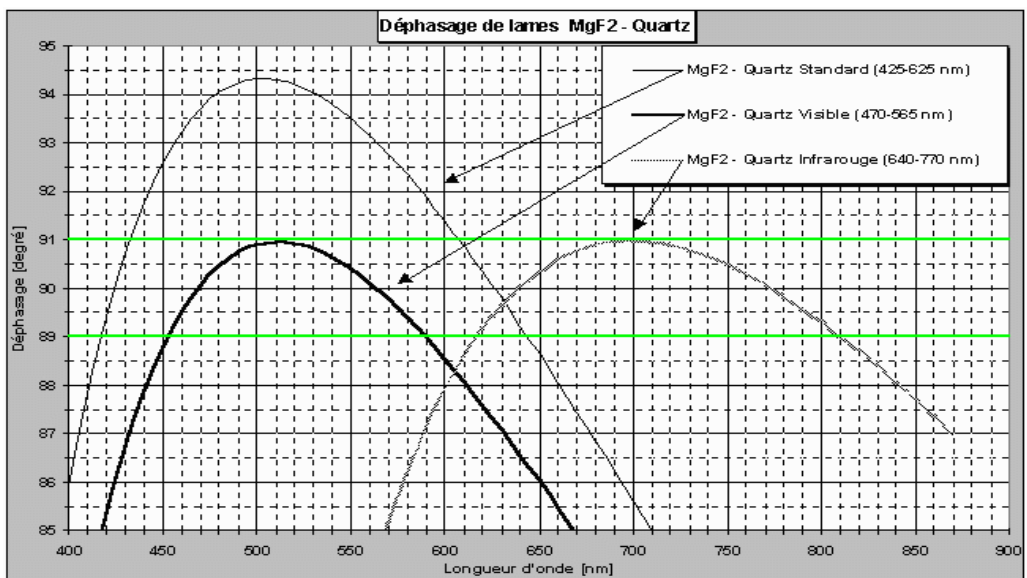

*Figure 3: Chromaticity of the so-called achromatic waveplates (assembly of quartz and MgF2) in the case of quarter waveplates of the "Fichou Company" as a function of the wavelength*

## I – 3 - The polarimetric grid

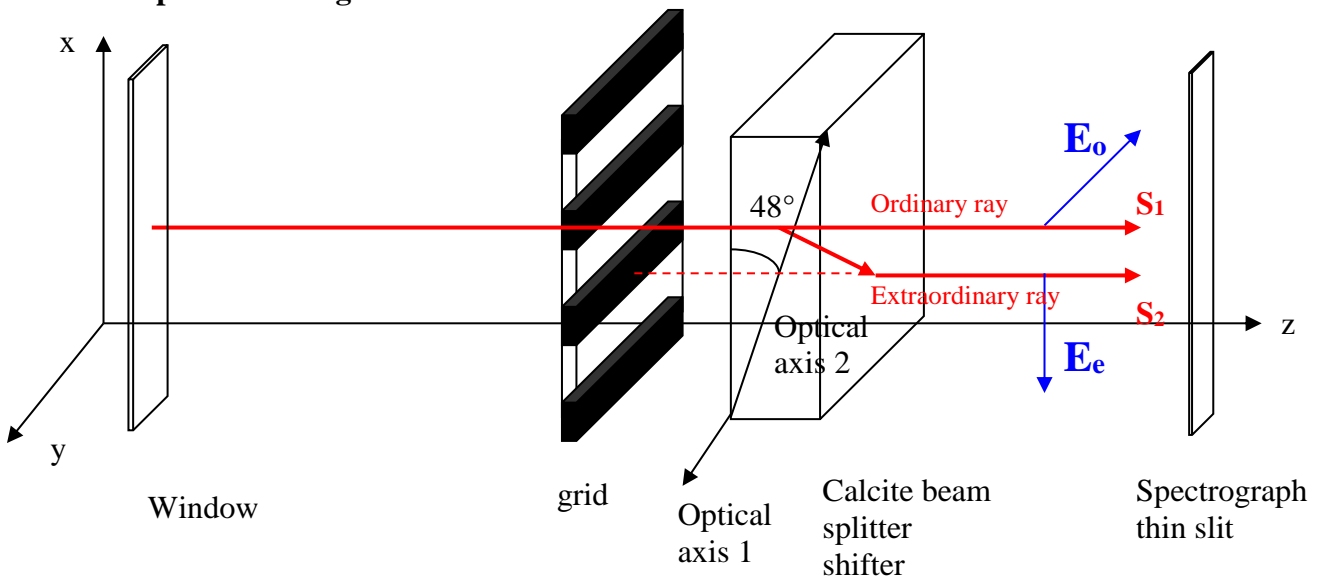

*Figure 4: the grid in the optical path*

The grid method was introduced by Semel (1980), it guaranties two optical paths as close as possible for the two beams, thus improving the polarimetric precision. The step of the grid must adapt to the thickness e of the separator so that the two beams $S_1$ and $S_2$ do not overlap (spatial shift of Figure 2). The grid and its effect are detailed in Figure 5.

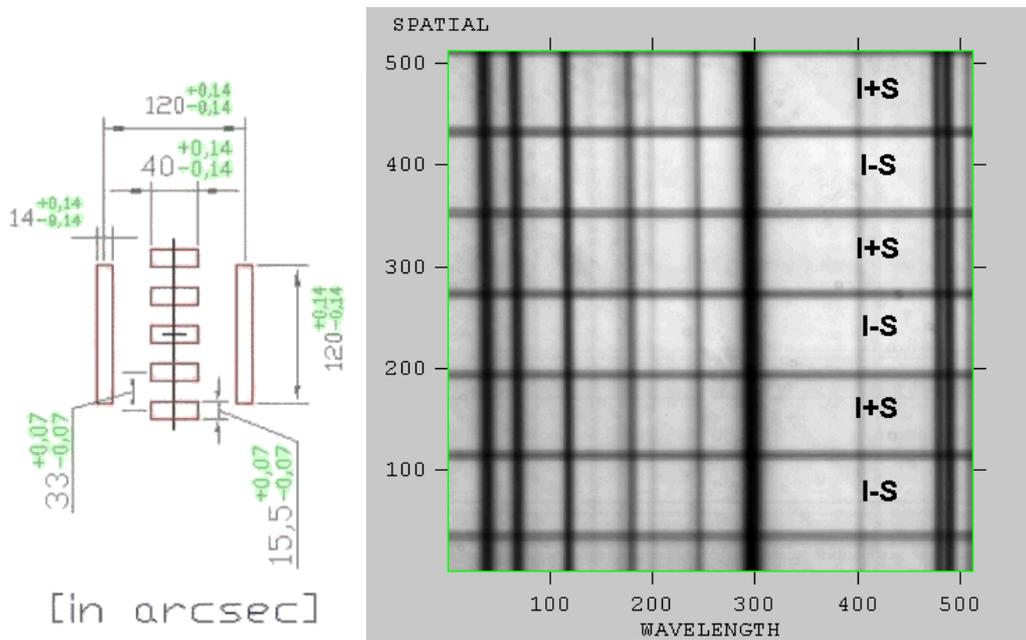

*Figure 5: the grid at the primary focus F1 (left) and its effect at the spectrum focus of the spectrograph (right). The height of the strips is 15". The grid allows to form I+S and I-S as simultaneous strips with S = (Q, U, V) in sequence, or S = (Q, -Q, U, -U, V, -V) in sequence with beam exchange. The FOV of the CCD camera (IXON 512 x 512) is 90" (6 x 15"). Courtesy OP.*

**I – 4 - Implementation of beam exchange on the Stokes vector measurement**

The beam exchange was introduced by Semel *et al* (1993). Let us consider Stokes Q polarimetry (it would be the same for V and U). the first measurement simultaneously gives $S_1 = I+Q$ and $S_2 = I-Q$ on two adjacent strips (Figure 5). The second measure gives $S'_1 = I-Q$ and $S'_2 = I+Q$ (beam exchange) on the same adjacent strips, the optical exchange (Q → -Q) is made by the quarter waveplates rotation. We suppose that these measurements (made at two successive times) remain strictly co-spatial. Let us introduce now $T_1$ and $T_2$, the transmissions of the two polarized beams. Hence, we measure sequentially:

At t1          at t2 (beam exchange)
$S_1 = T_1(I+Q)$    $S'_1 = T_1(I-Q)$
$S_2 = T_2(I-Q)$    $S'_2 = T_2(I+Q)$

Let us calculate the expression: $F = ¼ \ [ \ (S_1 S'_2) / (S'_1 S_2) \ - 1 \ ]$

we find that $Q/I = [ \ (4F+1)^{1/2} - 1] \ / \ [ \ (4F+1)^{1/2} + 1 \ ]$

When F <<1, the second order development provides $\boxed{Q/I = F - 2 F^2}$

This method does not require the knowledge of $T_1$ and $T_2$, i.e. the knowledge of the Flat Field and of the CCD gain table, which vanish in the F calculation. This is the interest of beam exchange, but it works only if the signals $S_1$, $S_2$, $S'_1$, $S'_2$ are co-spatial. In the case of variable seeing, the method loses interest because the signals obtained at time t2 are no longer co-spatial with those obtained at t1. So we mix information obtained at (x, y) at t1 with others obtained at (x+Δx, y+Δy) at t2, due to the atmospheric turbulence that agitates the images, produces differential distortions and defocuses. The method will only remain valid at low spatial resolution. Since the seeing evolves with a frequency of 100 Hz, the beam exchange method if efficient when a high-frequency polarimetric modulation is performed (1kHZ, in the case of ZIMPOL at IRSOL), or if there is an optical stabilization/correction of the seeing by a tip tilt or adaptive optics. At the epoch of our observations (2007), stabilization was not available, so that we may have seeing induced cross talk of the Stokes parameters, as the slit position may vary between t1 and t2, especially at the limb where there is a strong intensity gradient; however, tens or hundreds of accumulated exposures were done, so that most fluctuations should be smoothed in the summation process of the spectra. Also, THEMIS spectrograph is permanently rotating, and we noticed slow shifts of optical fringes and lines.

## II - THE DATASET (ZIP ARCHIVE)

- BaII4554
- BaII4932
- C2MgH5140
- CaI4227
- CaI6103
- Ha6563
- MgI5167
- MgI5173
- MgI5184
- ScII4247
- SrI4607
- Lande.txt
- Moore.txt

The dataset contains 11 directories of FITS files corresponding to 11 spectral domains, plus the Moore table and the Lande factors table (ASCII files).

The filenames of 32-bits FITS data files indicate:
- the location on the Sun (muXXX)
    DC = disk centre or $\mu = \cos\theta = 1.00$
    mu00, $\mu = \cos\theta = 0.00$ at the limb
    mu005, $\mu = \cos\theta = 0.05$ near the limb
    mu01, $\mu = \cos\theta = 0.10$ near the limb
    mu015, $\mu = \cos\theta = 0.15$ near the limb
- the sequence number (seqXX)
    Two observations with the same sequence number (in general one at DC and one value of $\mu = \cos\theta$ near the limb) were contiguous in time and were observed in similar conditions (in particular same exposure time, same spectral FOV). Hence, the DC observation may serve as a reference, it should have in most cases $Q = U = V \approx 0$ because it was done in the quiet Sun by moving the telescope around disk centre.

Each file contains a full header documented by the telescope (date, time, setup…) and a 3 dimensions FLOATING arrays; dimension 1 = wavelength; dimension 2 = slit direction; dimension 3 = Stokes parameter or polarization rate (I, U/I, Q/I, V/I). 3 horizontal strips of the polarimetric grid are present, their distance on the Sun is 15" in practice (45" total FOV segmented in 3 parts). Data were processed with "DeepStokes" available at THEMIS. The wavelength and abscissa along the slit are in pixels (no wavelength calibration). The solar atlas at disk centre (Delbouille's atlas JPEG files, showing also the Q/I polarization at $\mu = 0.1$ from Gandorfer's atlas) can be used for wavelength calibration using the disk centre (DC) files, which can also serve as Flat Field if needed. As there is almost no spatial resolution at the limb (because of tens or hundreds of accumulated files) or even at disk centre (because the Flat Field mode was used with the telescope describing "8" shaped loops), the 2D arrays can be summed along the slit direction (Y) to produce pure spectra and improve the S/N ratio. The slit (120" long) was always parallel to the limb.

## III - SOME RESULTS

CaI 4227 Å (Figure 6) shows unusually high Q/I values (7% at µ = 0, 6% at µ = 0.1, 3.5% at µ = 0.15), this is higher than previous determinations (3.3% at µ = 0, at Pic du Midi by Malherbe *et al*, 2007; 3.6% at µ = 0.1 by Bianda *et al*, 1998; 2.6% in Gandorfer's atlas, 2000, 2002, at µ = 0.1). This line is rather chromospheric than photospheric. At 423 nm, the waveplates of the polarimeter are no more exactly quarter wave (cos(86°) = 0.07 instead of cos(90°) = 0, this may introduce cross talk with other Stokes parameters or seeing induced cross talk (as we did not benefit of image stabilization).

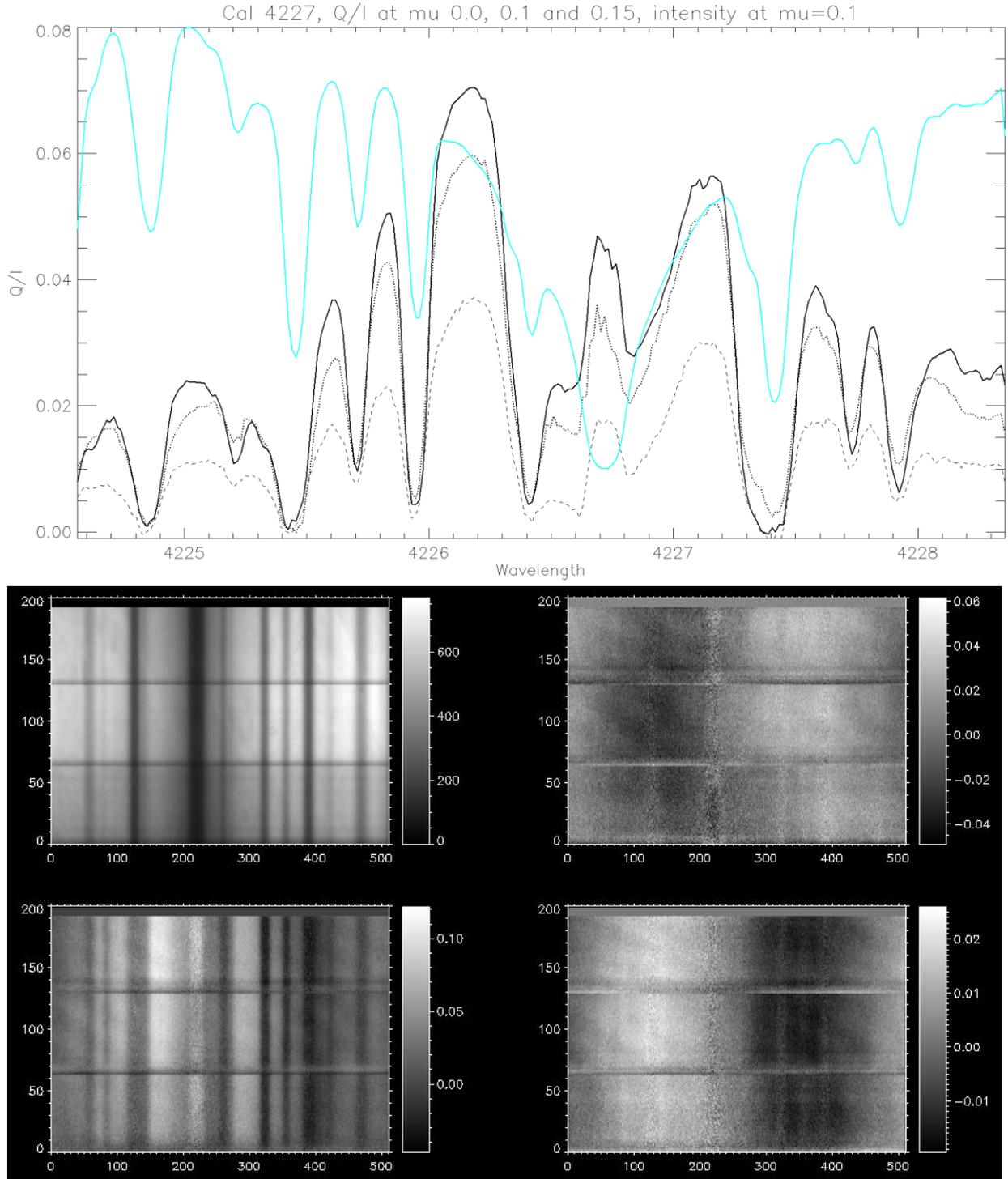

*Figure 6: CaI 4227Å. Top: Q/I (µ = 0, 0.1, 0.15) and I as a function of wavelength after integration along the slit. Bottom: four (λ, x) spectra of I (top left), Q/I (bottom left), U/I (top right) and V/I (bottom right); 3 strips of the grid (separated by 15") are visible. The wavelength (when indicated in pixels) varies from the right to the left.*

BaII 4554 Å (Figure 7) shows Q/I values of 1.4% at µ = 0, this is quite comparable to previous determinations (1.2% at µ = 0, Pic du Midi by Malherbe *et al*, 2007; 1.2% at µ = 0.1 by Stenflo & Keller, 1997). The hyperfine structure of Q/I (three peaks) is due to Barium isotopes; the central peak (even isotope) has 0 nuclear spin (so no hyperfine structure) while the secondary peaks (odd isotopes) have the 3/2 nuclear spin (hyperfine splitting) and 20% abundance.

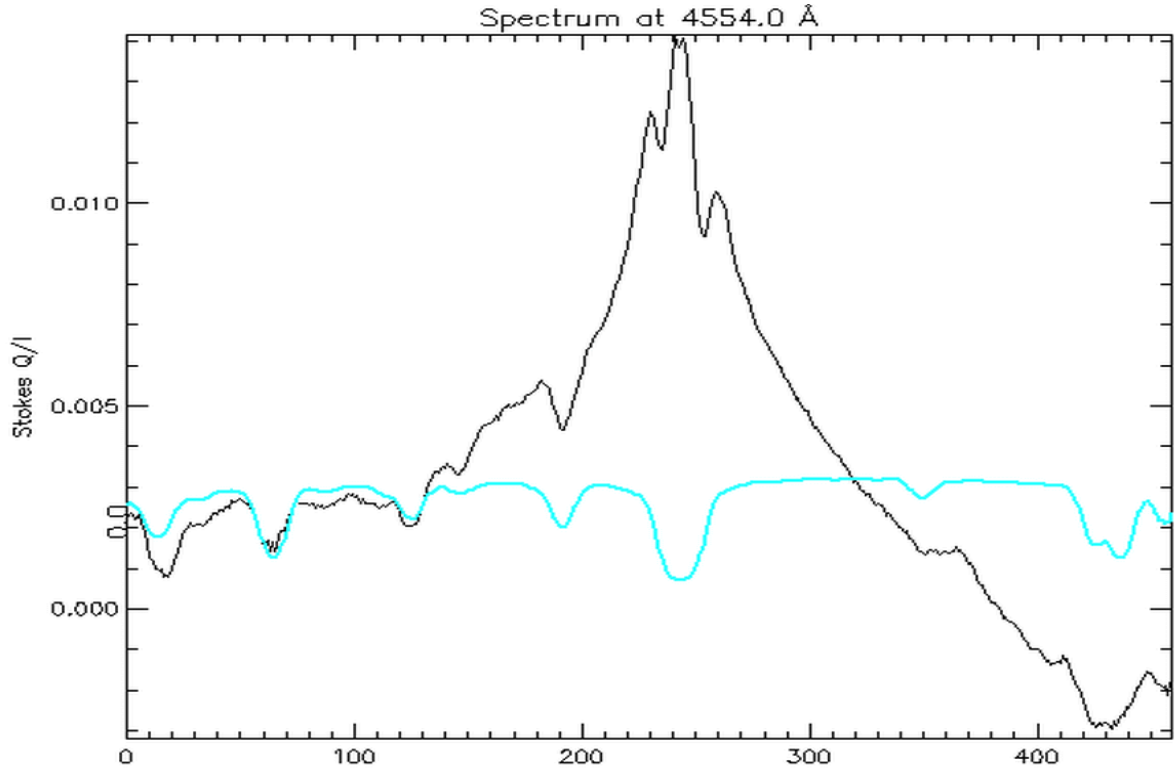

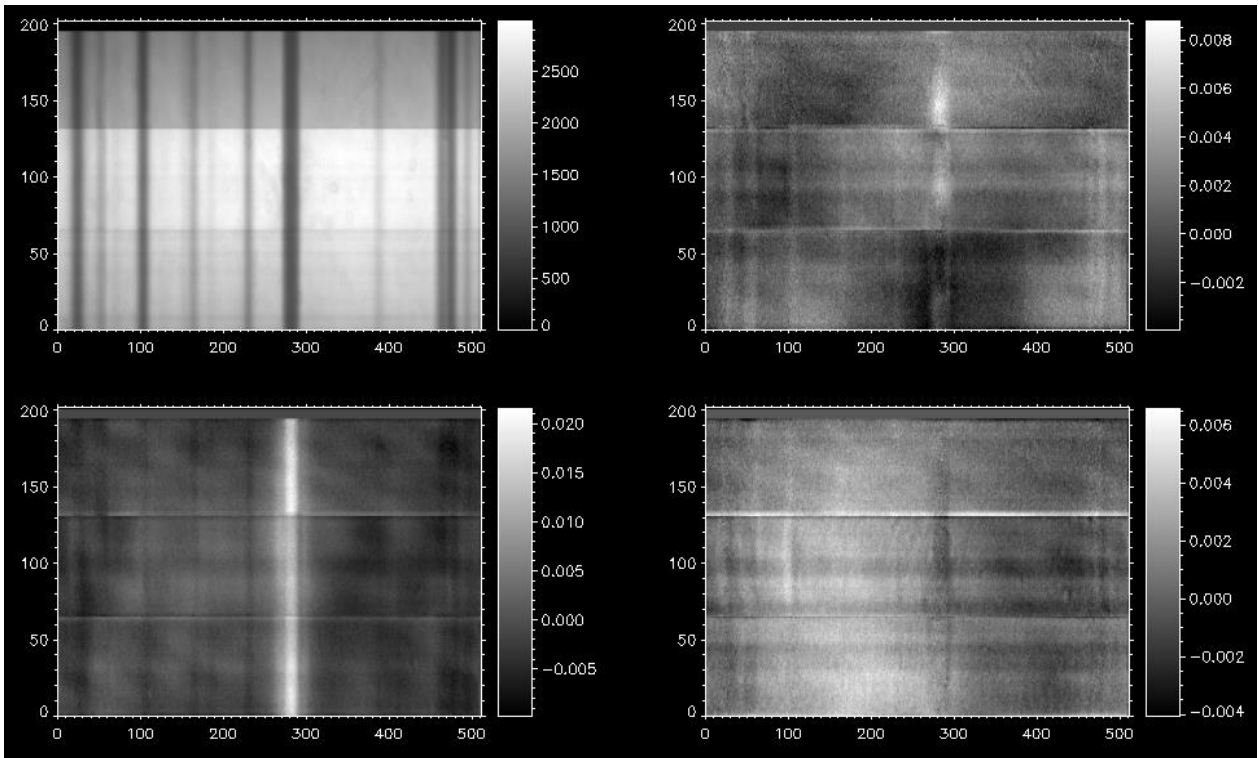

*Figure 7: BaII 4554 Å. Top: Q/I (µ = 0) and I as a function of wavelength after integration along the slit. Bottom: four (λ, x) spectra of I (top left), Q/I (bottom left), U/I (top right) and V/I (bottom right); 3 strips of the grid (separated by 15") are visible. Please notice the faint signal in U/I, this may indicate a rotation of the polarization direction (Hanle effect). The wavelength (when indicated in pixels) varies from the right to the left.*

SrI 4607 Å (Figure 8) shows Q/I values of 2.1% at µ = 0 and 1.6% at at µ = 0.1; this is comparable to previous determinations (1.8% at µ = 0, 1.2% at µ = 0.1, at Pic du Midi by Malherbe *et al*, 2007; 1.5% at µ = 0.1 by Stenflo & Keller, 1997).

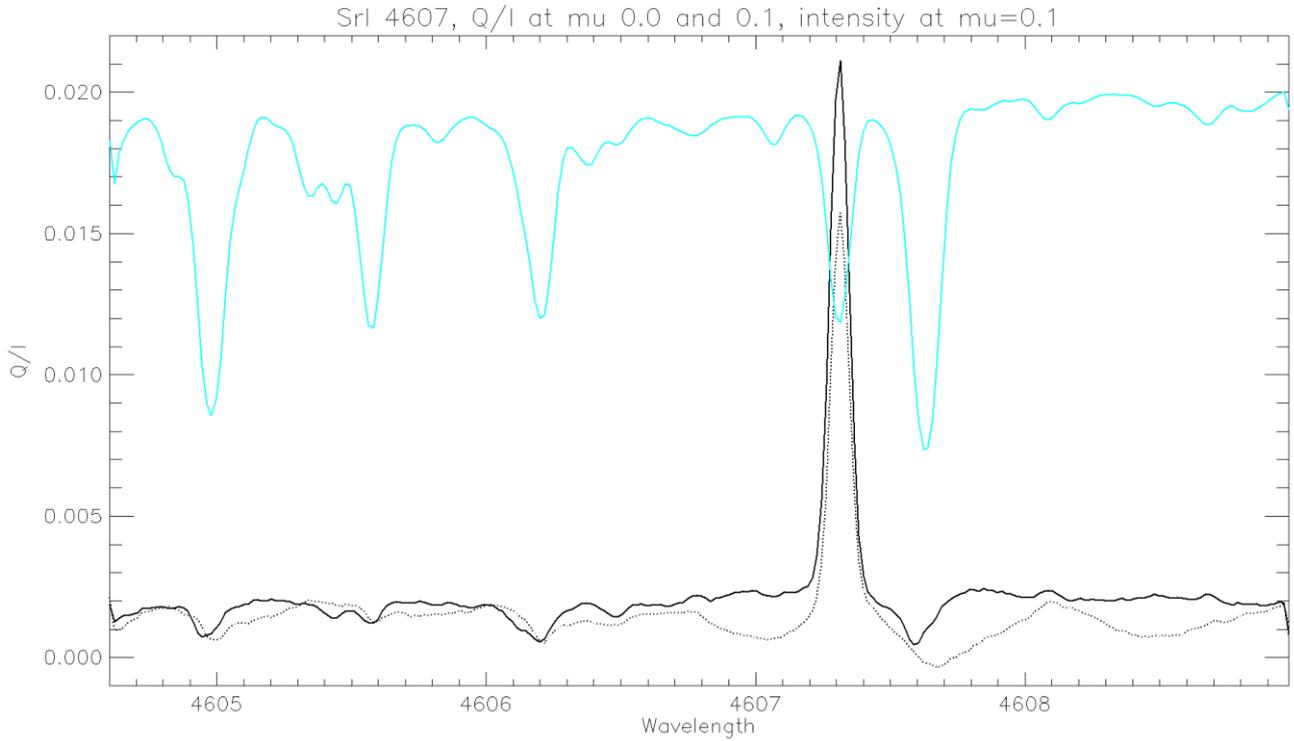

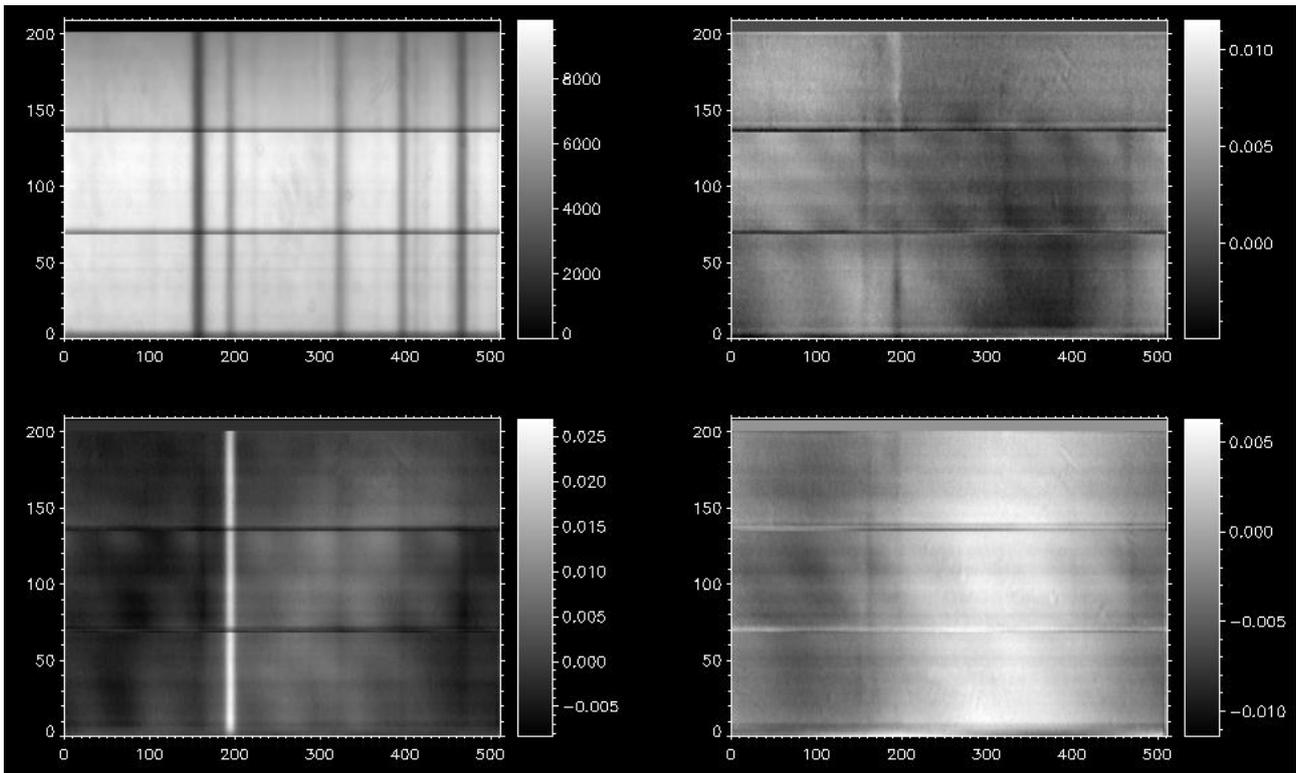

*Figure 8: SrI 4607 Å. Top: Q/I (µ = 0, 0.1) and I as a function of wavelength after integration along the slit. Bottom: four (λ, x) spectra of I (top left), Q/I (bottom left), U/I (top right) and V/I (bottom right); 3 strips (separated by 15") of the grid are visible. Please notice the faint signal in U/I, this may indicate a rotation of the polarization direction (Hanle effect). The wavelength (when indicated in pixels) varies from the right to the left.*

MgI 5184 Å (Figure 9) shows Q/I values of 0.18% at µ = 0; this is also comparable to previous studies (0.19% at µ = 0.1 by Stenflo *et al*, 2000).

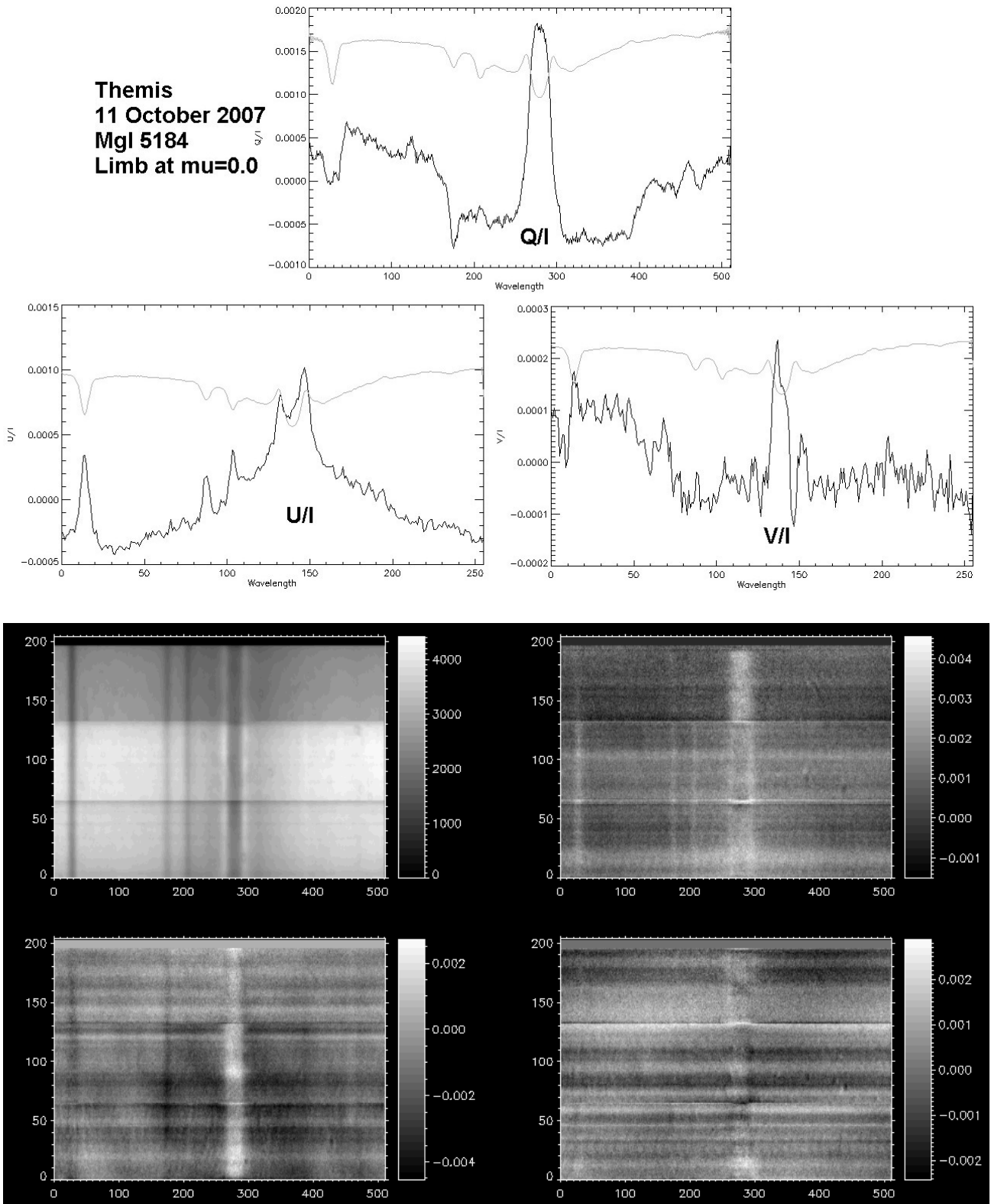

*Figure 9: MgI 5184 Å. Top: Q/I (µ = 0, 0.1) and I as a function of wavelength after integration along the slit. Bottom: four (λ, x) spectra of I (top left), Q/I (bottom left), U/I (top right) and V/I (bottom right); 3 strips (separated by 15") of the grid are visible. Please notice the faint signal in U/I, this may indicate a rotation of the polarization direction (Hanle effect). The wavelength (when indicated in pixels) varies from the right to the left.*

ScII 4247 Å (Figure 10) shows Q/I values of 0.4% at µ = 0; this is similar to previous measurements. The hyperfine structure of Q/I (three peaks) is due to the nuclear spin 7/2 implying splitting.

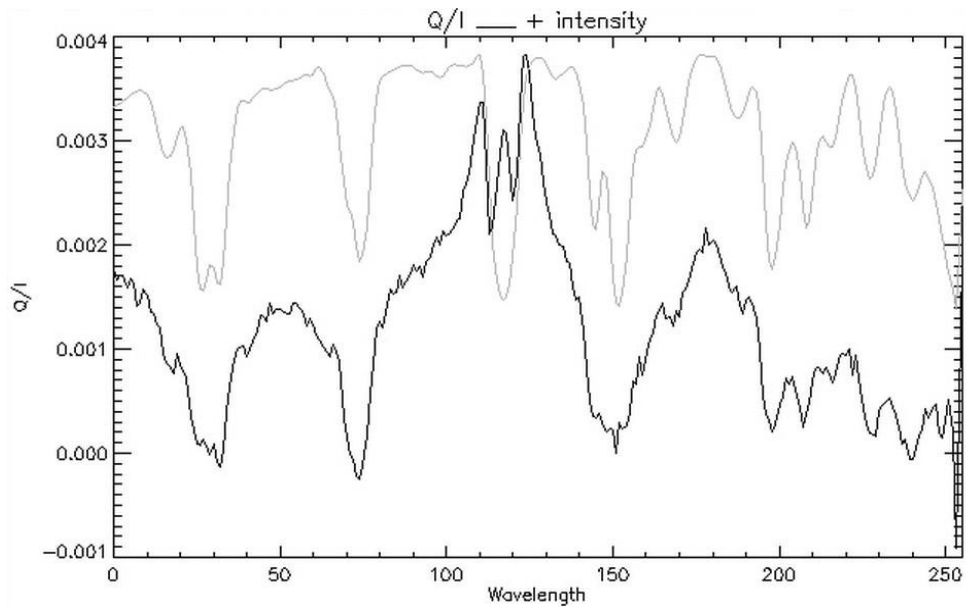

*Figure 10: ScI 5184 Å. Q/I (µ = 0) and I as a function of wavelength after integration along the slit. The wavelength (indicated in pixels) varies from the right to the left.*

**CONCLUSION**

Our measurements were performed as close as possible of the solar limb (µ ≈ 0) and in the vicinity of the limb (µ = 0.1, 0.15); most linear polarization rates Q/I at µ = 0.1 are in agreement with results already published. We find higher values at µ = 0 than at µ = 0.1, respectively 7%, 1.4% and 2.1% for CaI 4227 Å, BaII 4554 Å and SrI 4607 Å, this result is awaited. But the polarization Q/I of CaI 4227 Å is surprisingly higher than expected, this discrepancy could be due to cross talk as waveplates are no longer exactly quarter wave for this line, so that it is uncertain and new observations should be undertaken to corroborate this result.

**THE AUTHOR**

Dr Jean-Marie Malherbe (retired in 2023) is emeritus astronomer at Paris observatory. He first worked on solar filaments and prominences using multi-wavelength observations. He used the spectrographs of the Meudon Solar Tower, the Pic du Midi Turret Dome, the German Vacuum Tower Telescope, THEMIS (Tenerife) and developed polarimeters. He proposed models and MHD 2D numerical simulations for prominence formation. More recently, he worked on the quiet Sun, using satellites such as HINODE or IRIS, and MHD simulation results. He was responsible of the Meudon spectroheliograph from 1996 to 2023.